\documentclass[fleqn,usenatbib]{mnras}

\usepackage[T1]{fontenc}
\usepackage{ae,aecompl}

\usepackage{graphicx}	
\usepackage{amsmath}	
\usepackage{amssymb}	
\usepackage{bm}
\usepackage{float}

\newcommand{\deli}{\delta i}
\newcommand{\delj}{\delta j}
\newcommand{\dsub}{d_\text{sub}}
\newcommand{\Ssub}{S_\text{sub}}
\newcommand{\ensamble}[1]{\left\langle #1 \right\rangle}
\newcommand{\phase}[1]{\phi\left( #1 \right)}
\newcommand{\eref}[1]{Eq.(\ref{#1})}
\newcommand{\fref}[1]{Fig.\ref{#1}}
\newcommand{\tref}[1]{Table \ref{#1}}
\newcommand{\roverline}[1]{\mathpalette\doroverline{#1}}
\newcommand{\doroverline}[2]{\overline{#1#2}}
\newcommand{\Cobs}{\bm{C}_{\textbf{obs}}}
\newcommand{\Aobs}{\bm{A}_{\textbf{obs}}}
\newcommand{\Ctheo}{\bm{C}_{\textbf{theo}}}
\newcommand{\Ctheok}{\bm{C_{\textbf{theo},k}}}
\newcommand{\Atheok}{\bm{A_{\textbf{theo},k}}}
\newcommand{\Atheoi}{\bm{A}_{\textbf{theo},\infty}}

\title[Statistics of Turbulence at Maunakea]{Statistics of Turbulence Parameters at Maunakea using multiple wave-front sensor data of RAVEN}

\author[Y. H. Ono et al.]{
Yoshito H. Ono,$^{1,2}$\thanks{E-mail: yoshito.ono@lam.fr}
Carlos M. Correia,$^{1}$
Dave R. Andersen,$^{3}$
Olivier Lardi{\`e}re,$^{3}$\newauthor
Shin Oya,$^{4}$
Masayuki Akiyama,$^{2}$
Kate Jackson$^{6}$
and Colin Bradley$^{5}$\\
$^{1}$Aix Marseille Univ, CNRS, LAM, Laboratoire d'Astrophysique de Marseille, Marseille, France;\\
$^{2}$Astronomical Institute, Tohoku University, 6-3 Aramaki, Aoba-ku, Sendai 980-8578, Japan;\\
$^{3}$NRC-Herzberg, 5071 West Saanich Rd., Victoria, British Columbia, Canada;\\
$^{4}$TMT-J Project Office, NAOJ, 2-21-1 Osawa, Mitaka, Tokyo 181-8588, Japan;\\
$^{5}$Adaptive Optics Laboratory, University of Victoria, 3800 Finnerty Rd., Victoria V8P 5C2, British Columbia, Canada;\\
$^{6}$Division of Engineering and Applied Science, California Institute of Technology,\\1200 E. California Boulevard MC 155-44, Pasadena, CA 91125, USA;
}

\date{Accepted XXX. Received YYY; in original form ZZZ}

\pubyear{2016}

\begin{document}
\label{firstpage}
\pagerange{\pageref{firstpage}--\pageref{lastpage}}
\maketitle

\begin{abstract}
Prior statistical knowledge of the atmospheric turbulence is essential for designing, optimizing and evaluating tomographic adaptive optics systems. We present the statistics of the vertical profiles of $C_N^2$ and the outer scale at Maunakea estimated using a Slope Detection And Ranging (SLODAR) method from on-sky telemetry taken by RAVEN, which is a MOAO demonstrator in the Subaru telescope. In our SLODAR method, the profiles are estimated by a fit of the theoretical auto- and cross-correlation of measurements from multiple Shark-Haltmann wavefront sensors to the observed correlations via the non-linear Levenberg-Marquardt Algorithm (LMA), and the analytic derivatives of the spatial phase structure function with respect to its parameters for the LMA are also developed. The estimated profile has the median total seeing of 0.460$^{\prime\prime}$ and large $C_N^2$ fraction of the ground layer of 54.3\,\%. The $C_N^2$ profile has a good agreement with the result from literatures, except for the ground layer. The median value of the outer scale is 25.5\,m and the outer scale is larger at higher altitudes, and these trends of the outer scale are consistent with findings in literatures. 
\end{abstract}



\section{Introduction} \label{sec:intro}
Prior statistical knowledge of the atmospheric turbulence, such as layer altitude and stratified strength, outer scale, and wind velocity (speed and direction) is essential for designing and optimizing Adaptive-Optics (AO) systems in general and tomographic AO providing corrections over fields larger than the isoplanatic angle in particular.\par
Recently, Wide Field-AO (WFAO) systems have been developed for the current 8\,m-class telescope \citep{Neichel-14,Strobele-14,Vidal-14,Lardiere-14}, and are being designed for future Extreme Large Telescopes (ELT) \citep{Herriot-14,Thatte-14}, which have primary mirror diameters in the range 20--40\,m. Such WFAO systems require the vertical profile of the turbulence strength $C_N^2(h)$ to tomographically reconstruct the three-dimensional structure of the phase distortion caused by the atmospheric turbulence above the telescope.\par
Although parameters such as the coherence length ($r_0$ which is related to the seeing $s=0.98\lambda/r0$) and layer heights are relatively well constrained, the estimation of the outer-scale $\mathcal{L}_0$ with typical values of $\sim$20--30\,m at good observation sites has become an important research topic \citep{Ziad-04,Maire-07}. As we move towards larger apertures the impact of $\mathcal{L}_0$ on the estimation of seeing and $C_N^2$ becomes more important; moreover the vertical profile of $\mathcal{L}_0$ makes tilt angular decorrelation very different from constant $\mathcal{L}_0$ profiles, thus impacting estimation of tilt anisoplanatism and constraining system designs. The estimation of wind speed and direction $\bm{v}$ can be used in advanced temporal-control of AO systems \citep{Correia-14,Ono-16}. In addition, the knowledge of the atmospheric turbulence parameters is important for diagnostic and post-processing such the performance evaluation of AO correction and Point Spread Function (PSF) reconstruction.\par
Several techniques based on spatial or temporal correlation (or called as 'covariance') of the measured slope of Shack-Hartmann WFSs (SH-WFSs) were proposed to retrieve the vertical profiles of $C_N^2(h)$, $\mathcal{L}_0(h)$ and $\bm{v}(h)$ \citep{Wilson-02,Butterley-06,Cortes-12,Martin-16}, and already implemented into on-sky WFAO systems \citep{Vidal-14,Neichel-14,Lardiere-14}.\par
%
The SLOpe Detection And Ranging (SLODAR) is a method commonly used to estimate the vertical profile of $C_N^2$ in real time from measurements of multiple Shack-Hartmann WFSs (SH-WFS). This method is based on optical triangulation between two or more stars, and retrieves the vertical profile of $C_N^2$ from the intensities of peaks in the spatio cross-correlation deconvolved by the auto-correlation \citep{Wilson-02}. We refer to this method as the deconvolved-SLODAR. It's great advantage is that it is model-independent. In addition, using temporal cross-correlation of the measured slopes allows us to estimate the temporal features of the atmospheric turbulence \citep{Wang-08,Guesalaga-14}.\par
Another SLODAR approach is proposed in \cite{Butterley-06}, which recovers the vertical profile of $C_N^2(h)$ by fitting the theoretical spatio cross-correlation to the observed spatial cross-correlation. This second approach is referred to as the fitted-SLODAR in this paper. With the latter, unlike it's predecessor deconvolved-SLODAR, we'll be able to estimate the vertical profile of the $\mathcal{L}_0(h)$ by conforming to the von Karman power spectrum model for the theoretical spatio cross-correlation.\par
The SLODAR methods are thought to grasp in-situ from the real time AO telemetry more effects than off-site monitors that may look at different objects along different lines-of-sight not simultaneously with the AO observations. AO telemetry probes the atmosphere, the dome seeing and any instrument-specific aberrations which can be disentangled with appropriate processing.\par
In this paper we develop a new SLODAR method that minimises fitting residuals over the auto- and cross-correlation functions using analytic derivatives of the spatial phase structure function with respect to its parameters to accelerate the non-linear solver -- in our case the Levenberg-Marquardt Algorithm (LMA). We then post-process on-sky telemetry from multiple SH-WFSs installed on RAVEN, a Multi-Object AO (MOAO) technical and science demonstrator on the Subaru telescope. The first SLODAR result using RAVEN have been reported in \cite{Lardiere-14}. Here we improve our SLODAR method and provide consolidated statistics at Maunakea from a total of 12 nights May and August in 2014 and June and July in 2015. To our knowledge it is the first time that such achievement is reported for Maunakea based on the SLODAR method on 8\,m class telescope.
\par
This paper is organised as follows. In Section 2, we review a system of RAVEN, and on-sky observation of RAVEN on the Subaru telescope. Then, we present the theoretical model and non-linear minimization for our SLODAR in Section 3. In Section 4, we describe the estimated statistics of the vertical profiles of $C_N^2$ and $L_0$. Finally, we give some discussions in Section 5 and summarize our findings in Section 6.
%
%
%
%
\section{RAVEN} \label{sec:raven}%
RAVEN is a MOAO technical and science demonstrator on the Subaru telescope at Maunakea in Hawaii. RAVEN is the first MOAO demonstrator on 8\,m class telescope. The detail of the RAVEN system is summarised in \cite{Lardiere-14}. Here, we review shortly the system of RAVEN and our on-sky observations.\par
RAVEN applies MOAO corrections simultaneously into 2 science targets using a tomographic reconstruction. RAVEN has 4 open-loop (OL) SH-WFSs with 10$\times$10 subapertures for 3 natural guide stars (NGS) and 1 on-axis sodium laser guide star (LGS) installed in the Subaru telescope. Due to the obscuration by the telescope pupil, 72 subapertures in a SH-WFS are available, as shown in \fref{fig:pupil}.\par
The on-sky engineering and science observations with RAVEN have successfully been completed with the Subaru telescope. We have 12 nights in total: May and August in 2014 and June and July in 2015. One science paper based on RAVEN data has already been published \citep{Davidge-15} and other papers are coming soon. In total, 60\,hours on-sky telemetry from 3 OL-WFSs are recorded in the on-sky observations.\par
During the on-sky observations, we estimated the turbulence profile by a fitted-SLODAR with 3 NGS OL-WFSs for the tomographic reconstruction. The OL-WFSs were operated mostly with frame rates of 100--250\,Hz depending on brightness of natural guide stars. Brightness of natural guide stars is brighter than R<14\,mag in the most of the case. The spot position of each subaperture is measured mainly by the thresholded center of gravity (tCoG). We also tested the correlation centroiding method. The correlation centroiding provides slightly smaller measurement noise than tCoG for guide stars brighter than R<14\,mag, and, for guide stars
fainter than R>14\,mag, the correlation centroiding gives much smaller measurement noise \citep{Andersen-14}. From this fact, we expect that the measurement noise is negligible when the guide star magnitude is brighter than R<14\,mag because there is almost no difference in the centroiding accuracy of both the methods.\par
The SLODAR method used in the on-sky observations, which is referred to as the on-sky SLODAR, can measure only the vertical profiles of $C_N^2(h)$ and assumes an constant outer scale of 30\,m over all altitudes. In addition, the on-sky SLODAR measures the turbulence up to 12\,km to reduce the number of turbulence layers and to accelerate the computation of the tomographic reconstruction matrix. Although the on-sky SLODAR worked during the on-sky observations, this method is not enough to measure complete turbulence profile.\par
We also tried to estimate wind speed and direction at each altitude during on-sky observations using a method presented in \citep{Ono-16}, and tested new reconstruction algorithms \citep{Correia-15,Ono-16}, but in this paper we concentrate on retrieving the vertical profiles of $C_N^2(h)$ and $\mathcal{L}_0(h)$.
%
%
%
\section{SLODAR} \label{sec:method}
In this section we develop the SLODAR method improving upon the initial formulations to include the estimation of a $\mathcal{L}_0(h)$ profile. We start off by presenting how to compute the spatial correlations from WFS data and after show how to compute such functions analytically from von Karman spatial structure functions. The latter are then differentiated with respect to their parameters in order to compute the Jacobian as part of the non-linear optimisation routine that will minimise a $\chi^2$ criterion fitting data covariances to theoretical correlations. 
\begin{figure}
\centering
\includegraphics[viewport=0 0 377 376,width=.45\textwidth]{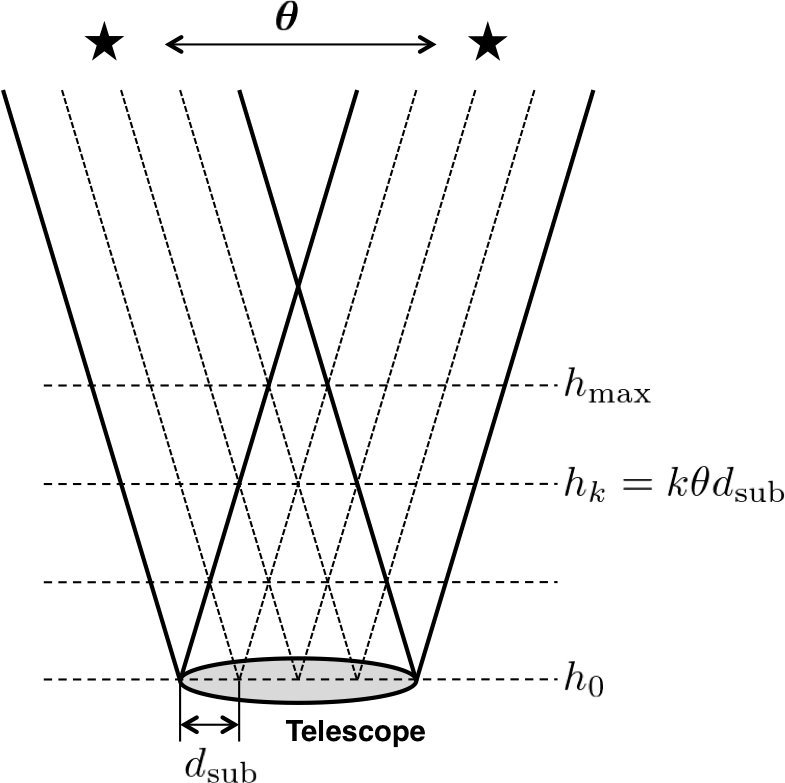}
\caption{Schematic image of SLODAR with two stars separating $\bm{\theta}$.} \label{fig:slodar}
\end{figure}
%
%
\subsection{Spatial correlations of slope data}
As mentioned previously, the SLODAR estimates the turbulence profile via the triangulation between the two stars, as shown \fref{fig:slodar}. A turbulence layer at altitude $h$ makes a peak in the spatio cross-correlation with a spatial offset corresponding to $h\theta$, where $\theta$ is an angular separation of the two stars. The offsets of the peaks in cross-correlation allow us to distinguish the atmospheric turbulence layers at different altitudes.\par
\begin{figure}
\centering
\includegraphics[viewport=0 0 576 432,width=.45\textwidth]{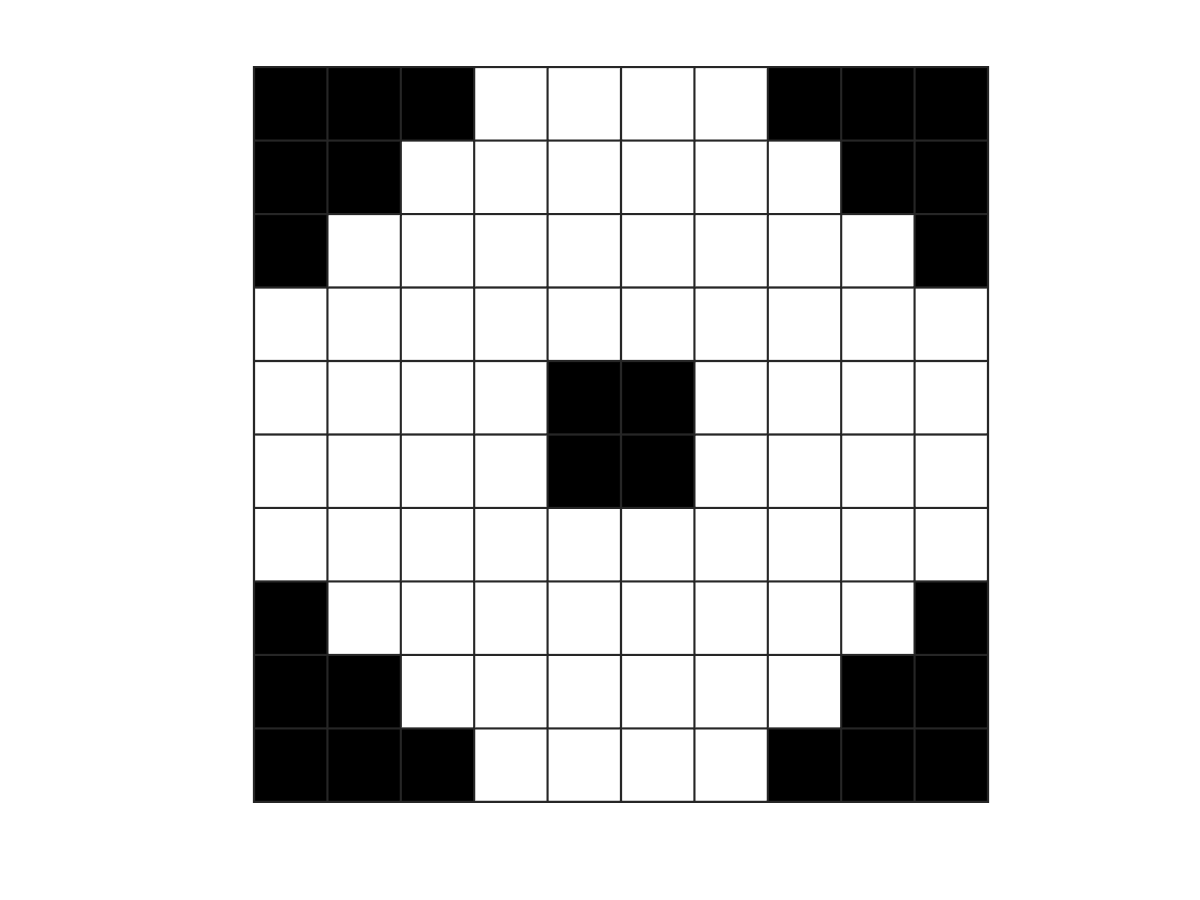}
\caption{Valid subapertures of SH-WFS in RAVEN.} \label{fig:pupil}
\end{figure}
The $x$-slope of the subaperture indexed as $(i,j)$ of $p$-th WFS is noted as $s^{x[p]}_{i,j}$. In order to remove the effect of overall tip/tilt caused by the telescope guiding error, wind-shake and vibration from telescope or/and instruments, the overall tip/tilt (i.e. mean slope over all subapertures) should be subtracted from each frame before the correlation is computed. The mean slope over all subapertures is given as
\begin{equation} \label{eq:averageslope}
\roverline{s^{x[p]}}=\frac{1}{N_\text{sub}}\sum_{i,j}s^{x[p]}_{i,j},
\end{equation}
where $N_\text{sub}$ is the total number of valid subaperture in a SH-WFS.
The spatio cross-correlation of $x$- and $x$-slopes between $p$-th and $q$-th WFSs, with the subaperture offset of $(\deli, \delj)$ and the tip/tilt removal, is defined as
\begin{equation} \label{eq:cross}
C^{x[p]x[q]}(\deli,\delj)=\frac{\sum_{i,j}\ensamble{\left(s^{x[p]}_{i,j}-\roverline{s^{x[p]}}\right)\left(s^{x[q]}_{i+\deli,j+\delj}-\roverline{s^{x[q]}}\right)}}{O(\deli,\delj)},
\end{equation}
where $\sum_{i,j}$ denotes a summation for all valid subapertures, $\ensamble{}$ denotes the average over the time series, $O(\deli,\delj)$ denotes the number of the valid subaperture pairs with the offset of $(\deli, \delj)$, which is equal to the spatio auto-correlation of a pupil mask shown in \fref{fig:pupil}. The auto-correlation for $p$-th WFS, $A^{x[p]}$, can be computed by \eref{eq:cross} taking $q=p$.\par
In \eref{eq:cross}, the cross-correlation is averaged for the same offset of $(\deli, \delj)$ by $O(\deli,\delj)$. This process makes the size of the correlation map small, and reduce the computational burden of SLODAR. In the case of RAVEN, the size of averaged and non-averaged correlation map is 19$\times$19 and 100$\times$100, respectively. This difference becomes critical for future extreme large telescopes, which have more than 5 WFSs and $\sim60\times60$ subapertures. Furthermore, the averaging process can make signal-to-noise ratio (SNR) high.\par
%
%
%
%
\subsection{Theoretical model of spatial correlation}
The SH-WFS measures a phase gradient averaged over a subaperture; it is modeled as
\begin{equation} \label{eq:slope}
s^{x[p]}_{i,j}=\int_{-\dsub/2}^{\dsub/2}\int_{-\dsub/2}^{\dsub/2} \frac{\partial \phase{x,y}}{\partial x}  dx dy/\Ssub,
\end{equation}
where we assume that all subapertures are squared, $(x,y)$ is a spatial coordinate with its origin at the center of subaperture, $\phase{x,y}$ is a phase value at $(x,y)$ and $\Ssub=\dsub^2$. Here, we use the approximated model for SH-WFS slope to accelerate the computation presented in \cite{Martin-12}. In the approximation the slopes are considered as the phase difference between two mid points on the both sides of the subapertures thus permitting the removal of the integrations from \eref{eq:slope} yielding
\begin{equation} \label{eq:slopeapprox}
s^{x[p]}_{i,j}\approx\frac{1}{\dsub}\left\{\phase{\frac{\dsub}{2},0}-\phase{-\frac{\dsub}{2},0}\right\}.
\end{equation}\ \par
We consider here a cross-correlation caused by single atmospheric layer at altitude $h$. In this case, the spatial distance $h\bm{\theta}$ between the projected pupils of two stars at altitude $h$ should be taken into account in the theoretical expression. We start from \eref{eq:slopeapprox} and use the definition of the phase structure function $D_\phi(\bm{\rho})=\ensamble{(\phi(\bm{r})-\phi(\bm{r}+\bm{\rho}))^2}$ and the equality  $2(A-a)(B-b)=-(A-B)^2+(A-b)^2+(a-B)^2-(a-b)^2$. With these definitions, the slope correlation corresponding to an atmospheric turbulence layer at altitude $h$ can be given as
\begin{align}\label{eq:theocorr}
\ensamble{s^{x[p]}_{i,j}s^{x[q]}_{i+\deli,j+\delj}}=&\frac{1}{2\dsub^2}
\left[-2D_\phi(\bm{\Delta_h})\right.\notag\\
&\left.+D_\phi(-\dsub\bm{u_x}+\bm{\Delta_h})+D_\phi(\dsub\bm{u_x}+\bm{\Delta_h})\right],
\end{align}
where $\bm{u_x}$ is a unit vector in $x$-direction, $\bm{\Delta_h}=(\Delta_x,\Delta_x)$ is projected distance at $h$ between the center of $(i,j)$ subaperture in $p$-th WFS and $(i+\deli,j+\delj)$ subaperture in $q$-th WFS, and $\Delta_x=\dsub\deli+h\theta_x$ and $\Delta_y=\dsub\delj+h\theta_y$. The spatial phase structure function for the von Karman power spectrum is given as
\begin{equation} \label{eq:vonkarman}
D_\phi(\rho)=0.17253\left(\frac{L_0}{r_0}\right)^{5/3}\left[1-\frac{2^{1/6}}{\Gamma(5/6)}\left(\frac{2\pi \rho}{L_0}\right)^{5/6}K_{5/6}\left(\frac{2\pi \rho}{L_0}\right)\right],
\end{equation}
where $K$ represents the modified Bessel function of the second kind. \par
The removal of overall tip/tilt should be also considered for the reasons pointed out before. The theoretical correlation with the tip/tilt removal is given as
\begin{align} \label{eq:subttcorr1}
&\ensamble{\left(s^{x[p]}_{i,j}-\roverline{s^{x[p]}}\right)\left(s^{x[q]}_{i+\deli,j+\delj}-\roverline{s^{x[q]}}\right)}\notag\\
&\ \ \ \ =\ensamble{s^{x[p]}_{i,j}s^{x[q]}_{i+\deli,j+\delj}}-\ensamble{\roverline{s^{x[p]}}s^{x[q]}_{i+\deli,j+\delj}}\notag\\
&\ \ \ \ \ \ \ \ -\ensamble{s^{x[p]}_{i,j}\roverline{s^{x[q]}}}+\ensamble{\roverline{s^{x[p]}}\hspace{1mm}\roverline{s^{x[q]}}},
\end{align}
where
\begin{align} \label{eq:subttcorr2}
\ensamble{\roverline{s^{x[p]}}\hspace{1mm}\roverline{s^{x[q]}}}&=\frac{1}{N_\text{sub}^2}\sum_{i',j'}\sum_{i,j}\ensamble{s^{x[p]}_{i',j'}s^{x[q]}_{i,j}},\\
\ensamble{\roverline{s^{x[p]}}s^{x[q]}_{i+\deli,j+\delj}}&=\frac{1}{N_\text{sub}}\sum_{i',j'}\ensamble{s^{x[p]}_{i',j'}s^{x[q]}_{i+\deli,j+\delj}}\label{eq:subttcorr3}
\end{align}
\eref{eq:subttcorr2} and \eref{eq:subttcorr3} show the auto-correlation of the mean slope and the cross-correlation between the mean slope and each slope, respectively.\par
All theoretical correlations in \eref{eq:subttcorr1}, \eref{eq:subttcorr2} and \eref{eq:subttcorr3} can be computed using \eref{eq:theocorr}. Then, the final expression of the tip/tilt removed theoretical correlation is given by substituting \eref{eq:subttcorr1} into \eref{eq:cross}. The $y$-$y$, $x$-$y$ and $y$-$x$ slope correlations can be given in a similar way, but in this work the $x$-$y$ and $y$-$x$ slope correlations are not used for turbulence profile estimation, because they have only weak correlation compared to the $x$-$x$ and $y$-$y$ slope correlations.\par
The theoretical model for the auto-correlation is given by a similar way to the cross-correlation as $q=p$. In the case of NGSs, the theoretical auto-correlation depend only on $C_N^2(h_k)$ and $L_0(h_k)$ and not on the altitude. On the other hand, in the case of LGSs, the theoretical auto-correlation depends on the altitude because the projected pupil size gets smaller with altitude due to the cone effect.
%
%
%
\subsection{Model fitting}
The vertical profile of $C_N^2(h)$ and $\mathcal{L}_0(h)$ can be recovered by fitting the theoretical correlations to the observed correlation. Here we define a vector $\bm{C}^{x[pq]}$ as a concatenation of $C^{x[p]x[q]}(\deli,\delj)$ for all subaperture offset $(\deli,\delj)$. The observed cross-correlation is noted as $\Cobs$, and the theoretical cross-correlation of $k$-th altitude bin is denoted as $\Ctheok$. When only the $C_N^2$ profile is estimated, the profile can be obtained by a linear fitting \citep{Butterley-06,Cortes-12}. However, a fit gets more complex when the $\mathcal{L}_0(h)$ are also estimated, because $\mathcal{L}_0(h)$ are non-linear parameters in the theoretical model.\par
\begin{figure}
\centering
\includegraphics[viewport=0 0 432 244,width=.45\textwidth]{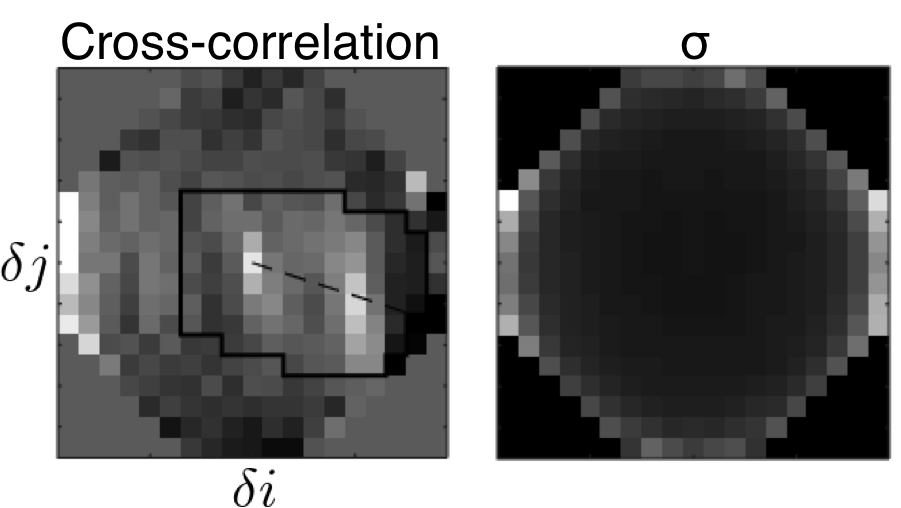}
\caption{Examples of the observed cross-correlation (left) and standard deviation of the cross-correlation (right) computed from the on-sky telemetry data taken by RAVEN. The dashed line in the left panel show the baseline of two stars. The area surrounded by the black line is used for the fitting.} \label{fig:example}
\end{figure}
The left panel of \fref{fig:example} shows an example of the observed cross-correlation computed from the 1 minute on-sky telemetry data taken by RAVEN. You can see the two correlation peaks on the baseline of 2 NGSs, shown as the dashed line. The central peak corresponds to the ground layer and the other peak corresponds to a high altitude layer. The edge of the cross-correlation map is very noisy due to the small number of subaperture pairs with large $(\delta i, \delta j)$. The similar trend can be seen in the standard deviation of the cross-correlation over 1 minute shown in the right panel of \fref{fig:example}. In order to reduce the effect from the noisy correlation values due to the small number of subaperture pairs, (1) the temporal standard deviation map is used as the weight of the fit, (2) the cross-correlation values with less than 5 subaperture pairs are removed from the fitting and (3) only the correlation values around the baseline are used in the fitting, which has a high signal-to-noise ratio, shown as the area in the black line in \fref{fig:example} \citep{Cortes-12}. It should be noted that, as the outer scale affects the shape of the correlation peak, the wide area around the baseline should be selected to estimate $\mathcal{L}_0(h)$ at each altitude.\par
The altitude is input in our SLODAR fitting. The altitude resoution $\delta h$ is given by $\delta h=d/\theta$ when the baseline of two stars is aligned to the $x$ or $y$-axis of a WFS lenslet array. On the other hand, when the baseline is not aligned to the $x$ or $y$-axis of the lenslet array, like \fref{fig:example}, the optimal resolution is given as $\delta h=d/(\theta\sin\alpha)$, where $\alpha$ is the angle of the baseline with respect to the $x$-axis for $\theta_x>\theta_y$ or the $y$-axis for $\theta_x<\theta_y$. In the case of RAVEN, the optimal $\delta h$ depends on the NGSs asterism and changes with time due to the field rotation, so that it should be updated during the observation. The maximum altitude $h_\text{max}$, which can be sensed via the triangulation using the cross-correlation, varies as well.\par
\begin{figure}
\centering
\includegraphics[width=0.45\textwidth]{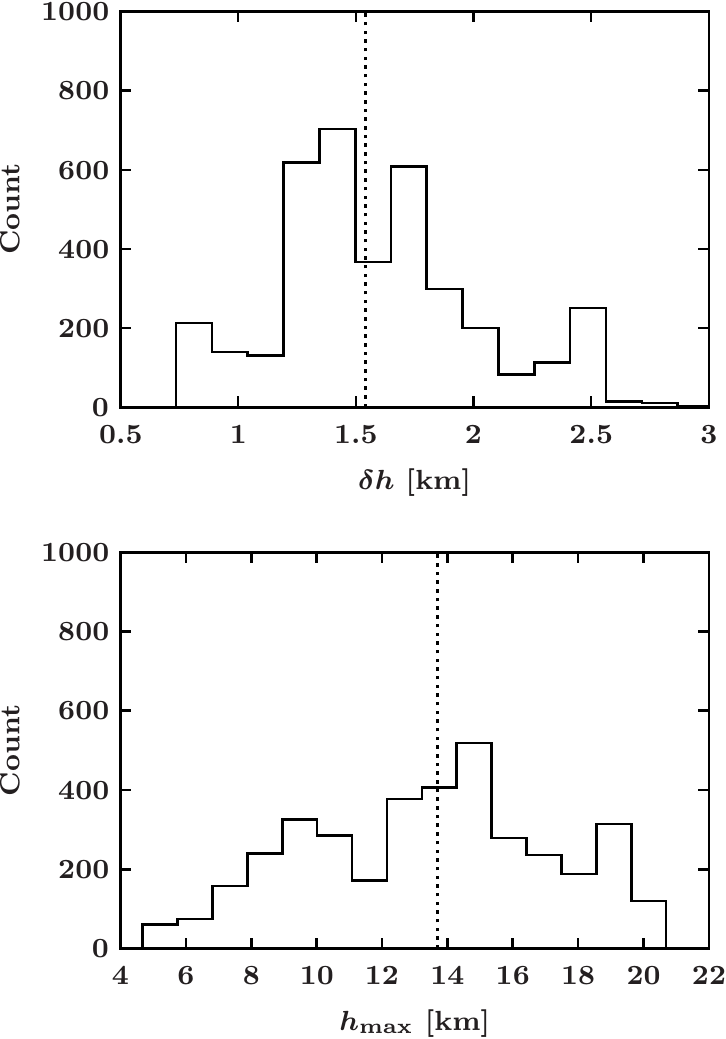}
\caption{Histograms of the altitude resolution $\delta h$ (top) and the maximum altitude $h_\text{max}$ (bottom) of SLODAR during the all on-sky observations. The vertical dotted lines show the median value, and it is 1.5\,km for $\delta h$ and 13.7\,km for $h_\text{max}$. The $\delta h$ and $h_\text{max}$ are scaled for the zenith direction.}
\label{fig:hmaxdh}
\end{figure}
Since RAVEN uses 3 NGSs, the SLODAR method can be computed with 3 different NGS pairs. Each pair has a different $\delta h$ and $h_\text{max}$, and the pair with narrow angular separation provides a small $\delta h$ and high $h_\text{max}$, whereas wide separation gives a large $\delta h$ and low $h_\text{max}$. In order to deal with all pairs together, we use the minimum altitude resolution in 3 pairs for all 3 NGS pairs. Although it causes the oversampling of altitude for the other 2 pairs with larger $\delta h$, we have more measurements from 3 pairs to derive the turbulence profile. The maximum altitude is determined by the GS pair with the smallest angular separation in the three pairs. \fref{fig:hmaxdh} shows histograms of $\delta h$ (the top panel) and $h_\text{max}$ (the bottom panel) during the on-sky observations. The range of the altitude resolution is from 0.75\,km to 3\,km and the median value is 1.5\,km. The maximum altitude ranges over a wide range and in some cases the turbulence only up to 5\,km can be sensed by the cross-correlation depending on the NGS configuration.\par
The turbulence above $h_\text{max}$ cannot be sensed via the triangulation with the cross-correlations (hereafter referred to as unsensed turbulence), but it can be measured by the auto-correlation of measurements from NGSs. In this paper, we use both the auto- and cross-correlations simultaneously to estimate the integrated $C_N^2$ and $\mathcal{L}_0$ of the unsensed turbulence. It should be noted that this method can not estimate the altitudes of the unsensed turbulences.\par
The $\chi^2$ value to be minimized in the fitting process is given as
\begin{align} \label{eq:cost}
\chi^2=&\left|\left|\ \sum_{p,q}\bm{W_{\rm c}}^{x[pq]}
\left[\Cobs^{x[pq]}-\sum_k^{N_\text{layer}}\Ctheok^{x[pq]}\right]\right.\right.\notag\\
&+\sum_{p,q}\bm{W_{\rm c}}^{y[pq]}
\left[\Cobs^{y[pq]}-\sum_k^{N_\text{layer}}\Ctheok^{y[pq]}\right]\notag\\
&+\sum_{p}\bm{W_{\rm a}}^{x[p]}
\left[\Aobs^{x[p]}-\left(\sum_k^{N_\text{layer}}\Atheok^{x[p]}+\Atheoi^{x[p]}\right)\right]\notag\\
&\left.\left.+\sum_{p}\bm{W_{\rm a}}^{y[p]}
\left[\Aobs^{y[p]}-\left(\sum_k^{N_\text{layer}}\Atheok^{y[p]}+\Atheoi^{y[p]}\right)\right]\ \right|\right|^2,
\end{align}
where $N_\text{layer}$ is the number of altitude bin and $\bm{W}$ represents the wighting matrix, which extracts the correlation values used in the fitting according to the criteria mentioned previously and a wight to each correlation value by a square inverse of the standard deviation of correlation over a time series. It is known that the central correlation value of the auto-correlation is enhanced by a correlation of measurement noise in slope \citep{Butterley-06}, and so we remove it from the fitting by $\bm{W_{\rm a}}$. The theoretical correlation is a function of altitude $h_k$, Fried parameter $r_{0,k}$ and outer scale $\mathcal{L}_{0,k}$, i.e. $\Ctheok=\Ctheo(h_k,r_{0,k},\mathcal{L}_{0,k})$, where $r_{0,k}$ relates to $C_N^2$ as
\begin{equation}\label{eq:r0}
r_{0,k}=\left[0.423k^2C_N^2(h_k)\right]^{-3/5},
\end{equation}
where $k=2\pi/\lambda$ and $\lambda$ is a wavelength.\par
We add the theoretical model of the unsensed turbulence, $\Atheoi$, only to the auto-correlation fitting term to estimate $C_N^2$ and $\mathcal{L}_0$ of the unsensed turbulence, which does not affect to the cross-correlation. So, the free parameters in \eref{eq:cost} are $r_0$ (or $C_N^2$) and $\mathcal{L}$ of $N_\text{layer}$ altitude bins and the unsensed turbulence; we have $2\times N_\text{layer}+2$ free parameters for the fitting.\par
\eref{eq:cost} is a non-linear least squares problem and can be simply expressed as $\chi^2=\sum_i[y_i-f(x_i,\bm{\beta})]^2/\sigma^2_i$, where $y$ is a measured value, $f$ is a model to be fitted, $\bm{\beta}$ represents parameters to be estimated and $1/\sigma^2$ is a weight. We use the Levenberg-Marquardt Algorithm (LMA) to determine the best parameters which minimizes $\chi^2$. The LMA find the best parameters iteratively, and in each iteration, the parameter $\bm{\beta}$ is updated to $\bm{\beta}+\bm{\delta\beta}$. In order to compute $\bm{\delta\beta}$, the model function $f$ is approximated as $f(x_i,\bm{\beta}+\bm{\delta\beta})\approx f(x_i,\bm{\beta})+\bm{J}\bm{\delta\beta}$,  where $\bm{J}$ is the Jacobian matrix and each element of $\bm{J}$ is a partial deviation of $f$ with respect to $\bm{\beta}$, thus $J_{ij}=\partial f(x_i,\bm{\beta})/\partial \beta_j$. In our case, $f$ consists of the theoretical correlations, and to compute the Jacobian matrix we need a partial deviation of the von Karman structure function in \eref{eq:vonkarman} with respect to $r_0$ and $\mathcal{L}_0$. In Appendix, we present how to compute this. In the iteration of the LMA, we add a condition that $r_0$ and $\mathcal{L}_0$ at all altitudes are lager than zero. In addition, an upper limit for $\mathcal{L}_0$ is set to 100\,m.\par
It should be noted that this method can be applied to a case in that altitudes of turbulence layers are free parameters, and also applied to the temporal correlation to estimate the wind speed and direction at each altitude as well. In the theoretical model, altitude and wind speed and direction affect the distance between 2 subapertures, this is, $\rho$ in \eref{eq:vonkarman}. For example, the partial deviation of \eref{eq:vonkarman} with respect to the altitude can be given as $\partial D_\phi/\partial h=(\partial \rho/\partial h)(\partial D_\phi/\partial \rho)$. The term of $\partial D_\phi/\partial \rho$ can be computed by a similar way to $\partial D_\phi/\partial \mathcal{L}_0$ shown in Appendix. The computation of $\partial \rho/\partial h$ depends on the model \citep{Martin-16}, but it is usually not complex.
%
%
%
\section{Results} \label{sec:results}
%
%
\subsection{Statistics of the atmospheric turbulence at Maunakea}
\begin{figure}
\centering
\includegraphics[width=0.45\textwidth]{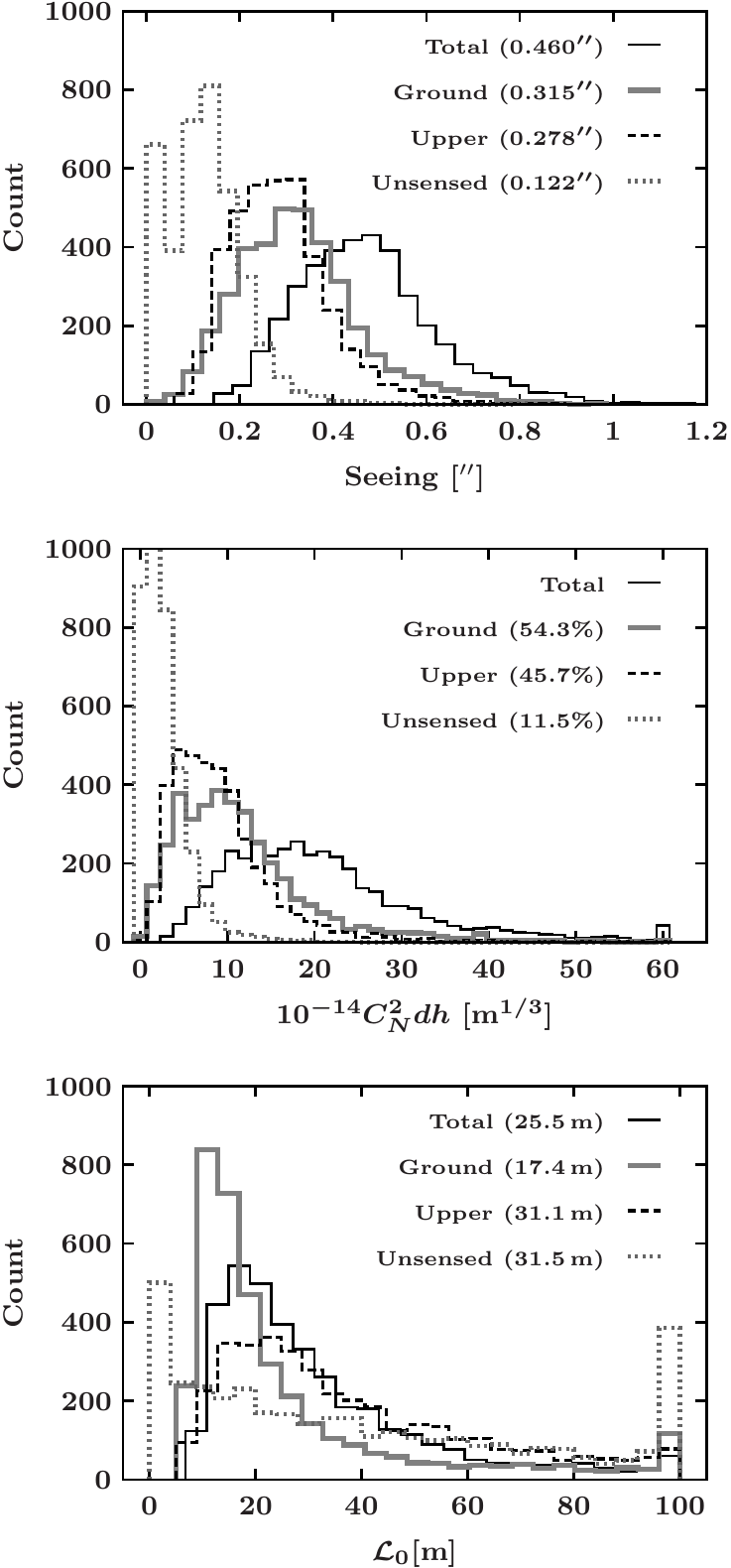}
\caption{Histograms of seeing (top), $C_N^2$ (middle) and $\mathcal{L}_0$ (bottom) for the total turbulence (black solid lines), the ground turbulence (gray bold lines), upper turbulence (black dashed lines) and the unsensed turbulence (gray dotted lines). The values in the parenthesises show median values for seeing, $C_N^2$ fraction and $\mathcal{L}_0$. The seeing is scaled for the zenith direction.} \label{fig:histogram}
\end{figure}
Here, we present the statistics of the atmospheric turbulence at Maunakea measured from the RAVEN on-sky telemetry data by our SLODAR. \fref{fig:histogram} shows histograms of seeing (top), $C_N^2$ (middle) and $L_0$ (bottom) for the total turbulence, the ground turbulence (0\,km$\leq h<1.5$\,km), the upper turbulence ($h\geq$1.5\,km, including the unsensed turbulence) and the unsensed turbulence. It is noted that our ground layer includes the turbulence up to 1.5\,km, and, therefore, it is not pure ground layer. The seeing is computed from the estimated $r_0$ via an equation of $\text{Seeing}=0.98\lambda/r_0$. The integrated $L_0$ over an altitude range is usually defined 
\begin{equation}
\mathcal{L}_0 (h_1<h<h_2)=\left(\frac{\int_{h_1}^{h_2}\mathcal{L}_0^{5/3}(h)C_N^2(h)dh}{\int_{h_1}^{h_2}C_N^2(h)dh}\right)^{3/5}.
\end{equation}
In this paper, seeing and $r_0$ are defined at $\lambda=$500\,nm. Also, these values are scaled for the zenith direction when it is not stated.\par
The median value of the total seeing during the RAVEN observations is 0.460$^{\prime\prime}$, and this is a quite good condition. The reason of this good seeing is that the on-sky observation was performed mostly in May, June and July in which the seeing gets small compared to other season \citep{Miyashita-04}.\par
\begin{figure}
\centering
\includegraphics[width=0.45\textwidth]{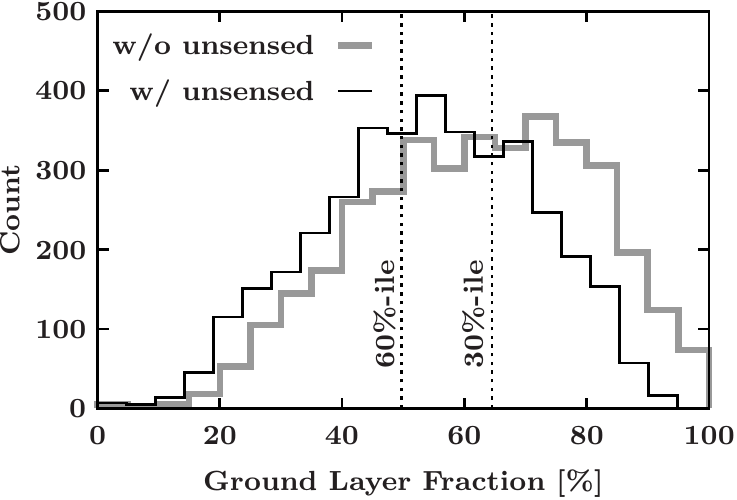}
\caption{Histogram of the $C_N^2$ fraction of the ground layer ($h<1.5$\,km). The black line shows the histogram including the unsensed turbulence and the vertical dotted line presents the ground layer fraction of 60\%-lie an 30\%-ile, which corresponds to the ground layer fraction of 50\% and 64.5\%, respectively. The gray bold line shows the histogram without the unsensed turbulence.}
\label{fig:glfrac}
\end{figure}

As expected, the ground layer has a large contribution and the median value of $C_N^2$ fraction of the ground layer is 54.3\%. Such a dominating ground layer at Maunekea has been reported in the previous results based on different methods: 85\% based on Multi-Aperture Scintillation Sensor (MASS) and Differential Image Motion Monitor (DIMM) fot Thirty Meter Telescope (TMT) site testing in \cite{Els-09}, and 54\% based on MASS-DIMM and 40\% based on SCIntillation Detection And Ranging (SCIDAR) in \cite{Tokovinin-05}. In addition, the large contribution by the ground layer can also be seen in \fref{fig:glfrac}, which shows a histogram of the $C_N^2$ fraction of the ground layer. From this figure the probability that more than 50\% of the turbulence is included in the ground layer (up to 1.5\,km) is 60\%.\par
\fref{fig:histogram} represents the existence of the unsensed turbulence above the maximum altitude that the SLODAR is sensitive to. In the middle panel of \fref{fig:histogram}, the $C_N^2$ histogram of the unsensed turbulence (shown in dotted gray line) has a narrow peak which can be seen to be a relatively small contribution with a median value of 11.5\% of the total. However, if we do not consider the unsensed turbulence, it is possible that the fraction of the ground layer is overestimated as shown in \fref{fig:glfrac}. We will come back to this discussion on the unsensed turbulence later in the paper.\par
The histograms of the outer scale have a steep slope at small $\mathcal{L}_0$ end and a long tail at large $\mathcal{L}_0$. This is because the SLODAR method is less sensitive to larger $\mathcal{L}_0$ than the telescope aperture, which affects the tip/tilt modes over the aperture, and we have a large uncertainty in large $\mathcal{L}_0$. The integrated outer scale is 25.5\,m in median, and it is consistent with a previous result at Maunakea (26.9\,m in \cite{Maire-07}). However, it is reported that $\mathcal{L}_0$ estimation by the SLODAR is biased to 2--3 times of the telescope aperture due to the less sensitiveness of SLODAR to large $\mathcal{L}_0$ \citep{Martin-16}, and our result is close to a size of 3 times of the telescope aperture. On the other hand, the similar or smaller values ($\mathcal{L}_0=$10--20\,m) have been found based on instruments with various spatial scales (1--100\,m) at a different site \citep{Ziad-04}. There are no definite results yet on the actual values of $L_0$ and more measurements with various instruments are required. The trend that the larger outer scale is at higher altitude in the histogram has been reported at several observation sites \citep{Maire-07,Guesalaga-16,Martin-16}. One possible explanation of this is since the SLODAR method is more sensitive to the large $\mathcal{L}_0$ at high altitudes thanks to the large meta-pupil at high altitudes.\par
%
\subsection{Median profiles}
\begin{figure}
\centering
\includegraphics[width=0.45\textwidth]{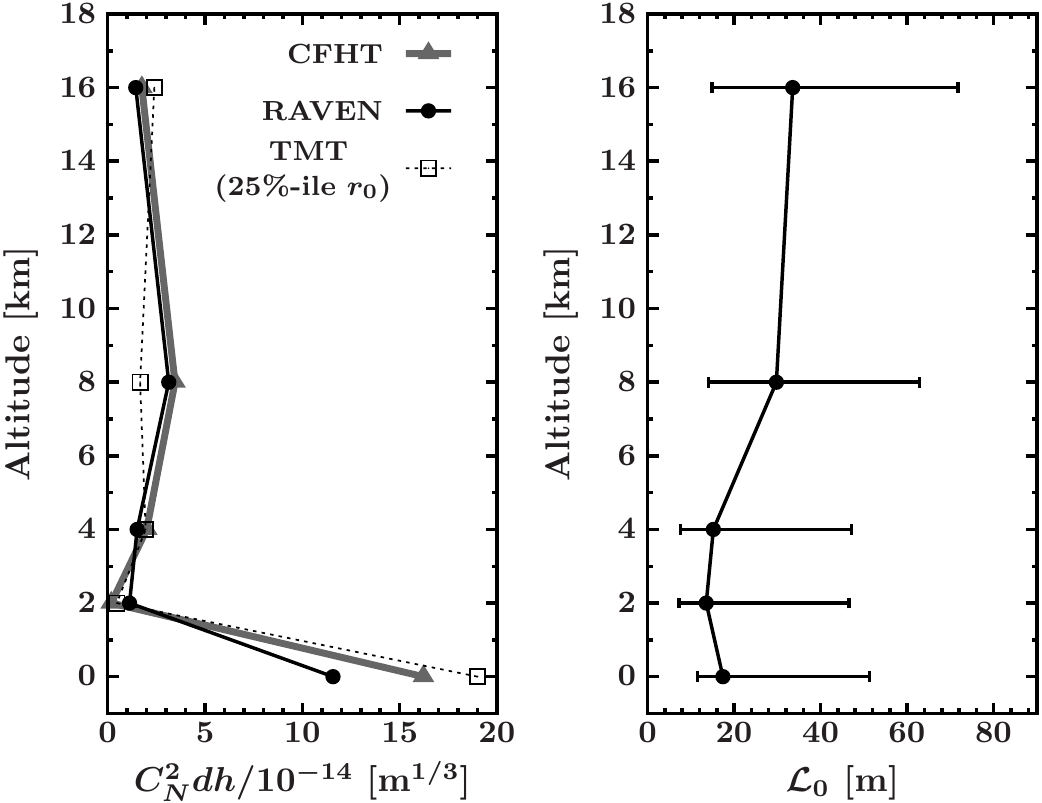}
\caption{Left: Mean $C_N^2$ vertical profile from the RAVEN SLODAR (black circles and solid line), the CFHT MASS and DIMM (gray triangles and bold line) and TMT site characterization with 25\%-ile $r_0$ (black squares and dotted line). Right: Median outer scale profile from the RAVEN SLODAR. The error bars are computed from standard deviations below and above the median $\mathcal{L}_0$. The values of the $C_N^2$ and the outer scale in the profiles are listed in \tref{tab:mean}} \label{fig:mean}
\end{figure}
\fref{fig:mean} and \tref{tab:mean} present the mean profile of $C_N^2$ and median profile of $\mathcal{L}_0$ noted as the black solid lines and circles. In order to compare our estimates with results from other instruments, the estimated profiles are resampled into 5 altitude bins: ground layer (0$\leq h\leq$1.5\,km), 2\,km (1.5$<h\leq$3\,km), 4\,km (3$<h\leq$6\,km), 8\,km (6$<h\leq$12\,km) and 16\,km (12$<h$). In the left panel of the figure, as mentioned previously, the ground layer has a strong contribution of $C_N^2$. There is a weak turbulence at 8\,km. At 2\,km, turbulence is not detected in most of the time.\par
As comparisons for the $C_N^2$ profile, we plot the mean $C_N^2$ profiles at Maunakea from a MASS and a DIMM at the Canada-France-Hawaii Telescope (CFHT; hereinafter, referred to as CFHT profile) and the site characterization over $\sim$2.5\,years for TMT in \cite{Els-09} (referred to as TMT profile). The DIMM at the CFHT monitored the total seeing in 9 nights out of 12 nights of the RAVEN observation. The MASS measured vertical profiles of $C_N^2$ at altitudes higher than 0.5\,km in 7 nights out of the our observations. The ground layer of the CFHT profile is estimated from a difference between a total seeing from the DIMM and the MASS. As the MASS and DIMM are not synchronized, the data overlapping each other within $\pm$60\,s are used for the ground layer.\par
The mean $C_N^2$ profile from the SLODAR has a good agreement with the CFHT profile, except for the ground layer, despite of the different locations of the Subaru and the CHFT telescopes. It means that the turbulence at high altitudes are relatively common over a wide range of sky, but the ground layer depends on the location. Also, the dome seeing of the Subaru also affects this difference in the ground layer. At 2\,km bin, the mean $C_N^2$ from the SLODAR is larger than the value of the CFHT. This is because the altitude resolution of the SLODAR is not enough to resolve the turbulence around the 2\,km bin and there is contaminated turbulence from other altitude bins through the resampling process.\par
For a TMT profile, we select a profile with good seeing conditions (25\%-ile $r_0$; seeing$<$0.55$^{\prime\prime}$), because our results correspond to the good seeing condition (the mean seeing is 0.46$^{\prime\prime}$). Similar to the comparison with the CFHT profile, there is a large difference in the ground layer between the RAVEN and the TMT profiles, and the weak ground layer in the RAVEN profile results in the good seeing condition during the RAVEN observation. The trend at high altitudes is different between the RAVEN and TMT profiles.\par
In the right panel of \fref{fig:mean}, the outer scale is larger at higher altitudes. The error bars of $\mathcal{L}_0$ show the standard deviations below and above the median $\mathcal{L}_0$, and the outer scale spread over a wide range as $\mathcal{L}_0$ is larger than the median value. These findings are consistent with the results in Section 4.1 and \fref{fig:histogram}.
\begin{table*}
\centering
\begin{tabular}{ccccccccc}
& & \multicolumn{3}{c}{Mean $C_N^2dh$ [$10^{-14}$m$^{1/3}$]} & \hspace{5mm} & \multicolumn{3}{c}{$\mathcal{L}_0$ [m]}\\[2pt]
\cline{3-5}\cline{7-9}\\[-4pt]
Altitude [km] &  range [km] & RAVEN & CFHT  & TMT   & & Median & $\sigma_\text{below}$  & $\sigma_\text{above}$ \\
\hline
\hline
0        &  0$\leq h\leq$1.5 & 11.58 & 16.22 & 19.00 & & 17.40 & 5.82  & 33.93 \\
2        &  1.5$<h\leq$3 &  1.12  & 0.19  & 0.48  & & 13.57 & 6.34  & 32.96 \\
4        &  3$<h\leq$6 &  1.52  & 2.00  & 1.95  & & 15.19 & 7.53  & 32.00 \\
8        &  6$<h\leq$12 &  3.14  & 3.46  & 1.67  & & 29.76 & 15.70 & 33.13 \\
16       &  12$<h$ &  1.45  & 1.75  & 2.40  & & 33.54 & 18.66 & 38.15 \\
\hline
\end{tabular}
\caption{Mean values of $C_N^2$ and median $\mathcal{L}_0$ at each altitude bin estimated by the RAVEN SLODAR. As comparison for the $C_N^2$ profile, mean $C_N^2$ profiles from the CFHT MASS-DIMM and the TMT site characterization in \protect\cite{Els-09} are also listed. For $\mathcal{L}_0$, $\sigma_\text{below}$ and $\sigma_{\rm above}$ represent the standard deviation computed below and above the median $\mathcal{L}_0$}
\label{tab:mean}
\end{table*}
%
%
\section{Discussion} \label{sec:discussion}
%
%
\subsection{Comparison with CFHT DIMM and MASS}
\begin{figure*}
\centering
\includegraphics[width=0.9\textwidth]{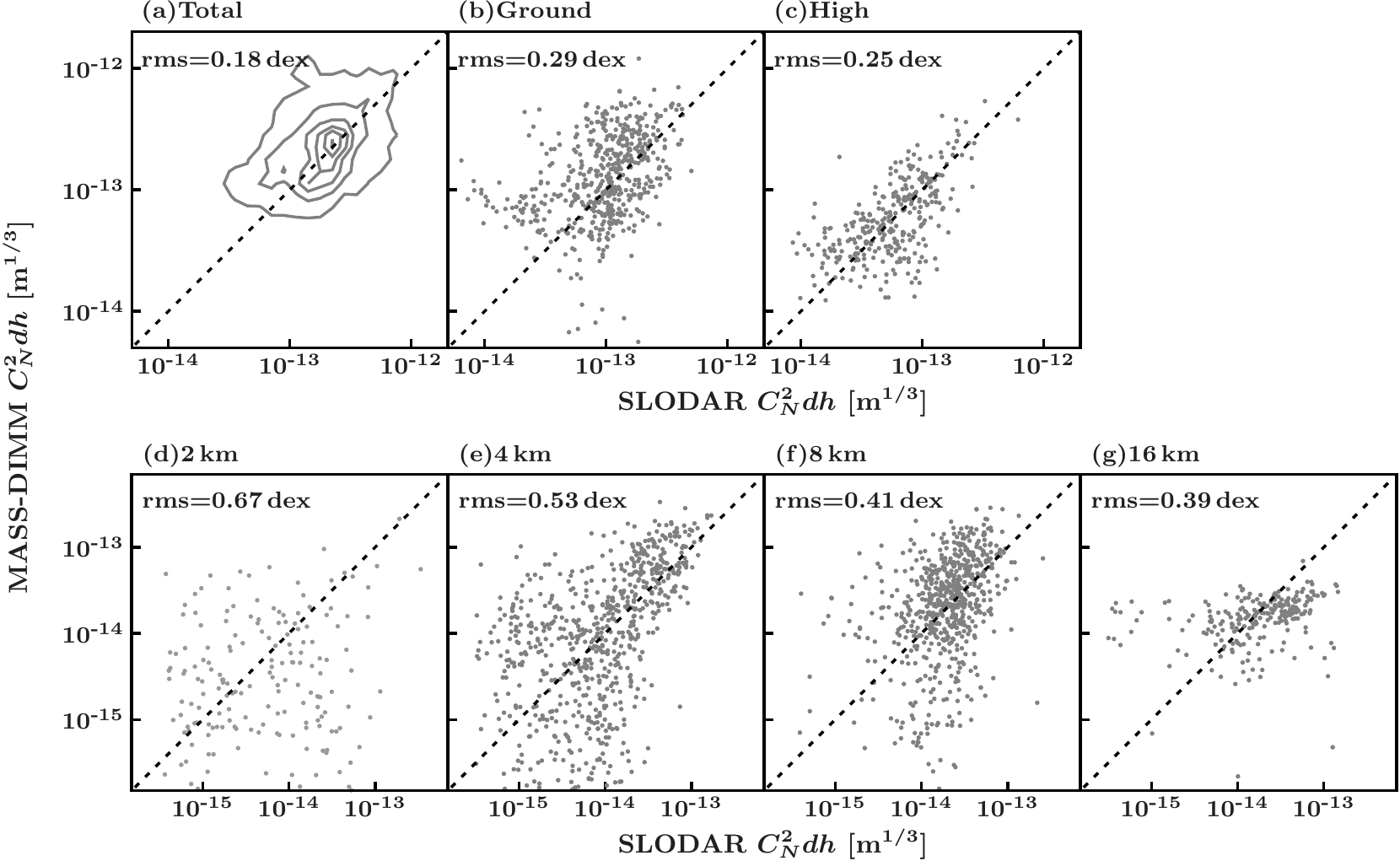}
\caption{Comparison of the estimated $C_N^2$ between the RAVEN SLODAR and the CFHT MASS-DIMM. (a) Total $C_N^2$ from the SLODAR and the DIMM. (b) Ground layer. For the CFHT profile, the ground layer is computed from the MASS and the DIMM data. (d)--(g) $C_N^2$ at 5 altitude bins used in Section 4.2. For the panel (a), we plot contours because the data points are too crowded to evaluate a correlation. In all panels, the turbulence weaker than $C_N^2dh=10^{-16}$\,m$^{1/3}$ is excluded because it is too weak to compare. The scatter rms value from the 1 to 1 relation (black dashed lines) are presented at upper right of each panel in units of dex.}
\label{fig:compare}
\end{figure*}
In Section 4, we compare the mean $C_N^2$ profile estimated by our SLODAR with that from the CFHT MASS-DIMM, and there is a good general agreement with each other except for the ground layer. In this section, we discuss this comparison in more detail. However, it should be noted that the CFHT MASS-DIMM cannot be perfectly compared to our SLODAR because these instruments have different altitude resolutions and observe different directions; moreover, Subaru and CFHT are located at at different places atop Maunakea.\par
\fref{fig:compare} shows the comparison of $C_N^2$ from the SLODAR and the MASS-DIMM at each altitude bin used in \fref{fig:mean}. The comparison for the total $C_N^2$ and $C_N^2$ at high altitudes (h$>$1.5\,km) are also shown in the figure. In \fref{fig:compare}, the turbulence weaker than $C_N^2dh=10^{-16}$\,m$^{1/3}$ is not plotted because such a weak turbulence is affected by the measurement noise and difficult to compare.\par
In the panel (a), the total $C_N^2$ estimated by the SLODAR correlates with the total $C_N^2$ estimated by the DIMM. In the panels of (b), the ground layer also correlates with each other, but having larger scatter (rms=0.29\,dex) than that of the total $C_N^2$ relation (rms=0.18\,dex) in the panel (a). The larger scatter of the ground turbulence may be affected by the contamination from other altitude bins due to not enough SLODAR altitude resolution as mentioned in Section 4.2. The ground layer in the CFHT profile tend to be slightly larger than the values in the SLODAR as shown in the mean $C_N^2$ values in Section 4.1. In addition, the ground layer is affected by the dome seeing.\par
The $C_N^2$ values at the high altitude in the panel (c) show a good agreement. However, $C_N^2$ relation between the SLODAR and the MASS at each altitude bin, shown in the panels (d)--(g), shows worth correlation compared to the correlation of the all high altitudes in the panel (c). One reason of less correlation at each altitude bin is the contamination from other altitude bins and this has a large impact at 2\,km in the panel (d). Also, at 2\,km, the turbulences weaker than $C_N^2=10^{-16}$\,m$^{1/3}$ are mostly detected, which is not included in the figure. At 4\,km and 8\,km, we can see good correlations between the $C_N^2$ from the SLODAR and the MASS as $C_N^2>10^{-14}$\,m$^{1/3}$, but there are a large dispersion at $C_N^2<10^{-14}$\,m$^{1/3}$ due to the contamination from other altitude bins. If we exclude the turbulence weaker than $C_N^2<10^{-14}$ from the rms computation, the scatter rms values are 0.25\,dex and 0.26\,dex for 4 and 8\,km bin, respectively, which is much smaller than the rms values including the all data points. In the highest altitude bin at 16\,km, the scatter rms of $C_N^2$ is smaller than the values for 4 and 8\,km, but the number of the data points is small because the maximum altitude of the SLODAR is limited below 12\,km depending on the guide star configuration.\par
Considering a large difference at the ground layer between the SLODAR and the MASS-DIMM, the turbulence profile for the tomography in WFAO systems should be estimated from its WFSs, not from a different system at a different place like MASS and DIMM. Otherwise, we would mis-estimate the ground layer, which has a large contribution of more than 50\% of total $C_N^2$ at Maunakea, and it has a large impact on the performance of WFAO systems.
%
%
\subsection{Is outer scale biased?}
\begin{figure}
\centering
\includegraphics[width=0.45\textwidth]{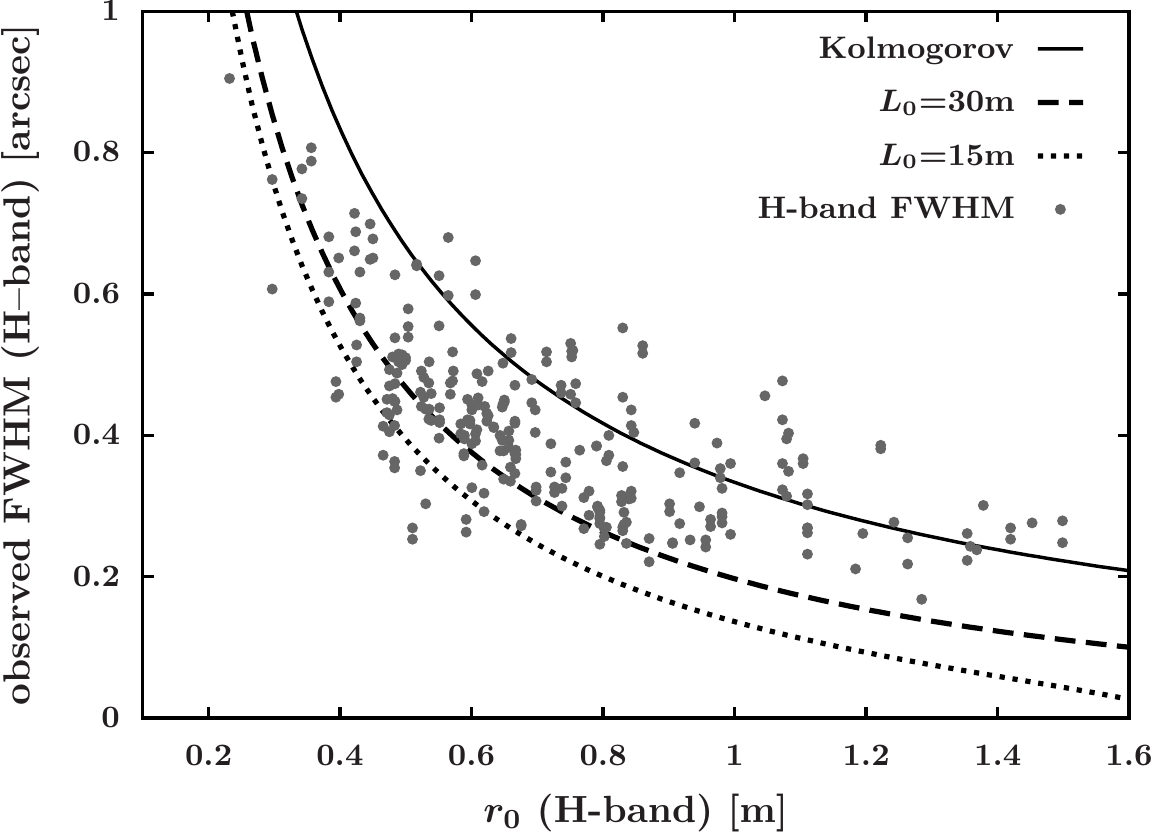}
\caption{Seeing-limited FWHMs of H-band PSFs as a function of $r_0$ at H-band. The seeing-limited FWHMs observed by RAVEN are represented by gray points. The black lines are the relation predicted from \eref{eq:seeingL0} with different $\mathcal{L}_0$: $\mathcal{L}_0=\infty$ (Kolmogorov; solid line), $\mathcal{L}_0=30$\,m (dashed line) and $\mathcal{L}_0=15$)\,m (dotted line).} \label{fig:noaofwhm}
\end{figure}
The FWHMs of PSFs without AO correction, referred as seeing-limited FWHM, are affected by the outer scale because the amount of tip/tilt components depends on the outer scale. In other words, the outer scale can be evaluated from the observed seeing-limited FWHMs by comparing it with those assuming the Kolmogorov power spectra.\par
In \cite{Tokovinin-02}, the approximation of the ratio of the seeing-limited FWHM assuming the von Karman power spectrum to that assuming the Kolmogorov power spectra is investigated though a numerical simulation and it is given as
\begin{equation}\label{eq:seeing}
\left(\frac{\epsilon_\text{vK}}{\epsilon_\text{Kol}}\right)^2\approx
1-2.183\left(\frac{r_0}{\mathcal{L}_0}\right)^{0.356},
\end{equation}
where $\epsilon_\text{vK}$ and $\epsilon_\text{Kol}$ are seeing-limited FWHMs at a given wavelength $\lambda$ assuming von Karman and Kolmogorov power spectrum, respectively. Using \eref{eq:seeing} and $\epsilon_\text{Kol}=0.98\lambda/r_0$, we can get a relation between $\epsilon_{\text{vK},\lambda}$ and $r_{0,\lambda}$ at a given wavelength $\lambda$ as
\begin{equation}\label{eq:seeingL0}
\epsilon_{\text{vK},\lambda}=0.98\frac{\lambda}{r_{0,\lambda}}
\sqrt{1-2.183\left(\frac{r_{0,\lambda}}{\mathcal{L}_0}\right)^{0.356}},
\end{equation}
where $r_{0,\lambda}$ can be computed from the $r_0$ at 500\,nm measured by the SLODAR using a relation $r_0\propto \lambda^{1.2}$.\par
We compare the seeing-limited FWHMs of H-band PSFs observed by RAVEN with those predicted from $r_{0,\lambda}$ estimated by the SLODAR using \eref{eq:seeingL0} to evaluate $\mathcal{L}_0$ in \fref{fig:noaofwhm}. The gray points in the figure show the seeing-limited FWHMs measured from on-sky PSF at H-band by a fit of elliptical Moffat function. We plot the FWHM in minor-axis to minimize the effect from tip/tilt induced by the telescope guiding error, wind-shake and vibration. The back lines show the prediction from \eref{eq:seeingL0} with different outer scales. According to \fref{fig:noaofwhm}, most of the on-sky seeing-limited FWHMs have a good agreement with the prediction with $\mathcal{L}_0>30$\,m. This value is larger than the median value of 25.5\,m estimated by the SLODAR, and it indicates that the estimates of the outer scale from the SLODAR is possibly biased to 2--3 times of the telescope aperture. \par
The outer scale has a large impact on the turbulence strength, especially on the strength of the low order modes, and so, the outer scale strongly affects designing AO systems such as dynamical ranges of DMs and WFSs and predictions of AO performance based on numerical simulations. Therefore, further measurements of $\mathcal{L}_0$ in different methods are required to estimate the actual outer scale for designing future AO systems at Maunakea.\par
The other thing that should be discussed is an impact from the bias effect of the outer scale in terms of tomography for WFAO systems. As mentioned previously, the outer scale affects mainly the tip/tilt modes of the phase aberration. The tomography method, which controls the low- and high-order aberration separately, is proposed in \cite{Gilles-08}, and this method can help to reduce the effect from the biased outer scale. At high altitudes, due to larger meta-pupil size in the atmospheric turbulence volume, the outer scale may affect the higher-order of the phase distortion than the tip/tilt modes and cause different tilt anisoplanatism over the field. However, if the SLODAR has an ability to sense larger outer scale at high altitudes thanks to the larger meta-pupil, the outer scale effect at high altitudes can be taken into account in tomography. \par
In the case of ELTs with a primary aperture larger than 30\,m, we will be able to sense the outer scale roughly up to 100\,m by the SLODAR method, which is much larger than the typical outer scale (20--30\,m) observed at some sites. Therefore, the impact from the bias effect in the SLODAR gets much smaller compared to the cases with current 8\,m class telescopes.
%
%
\subsection{Unsensed turbulence}
\begin{figure}
\centering
\includegraphics[width=0.45\textwidth]{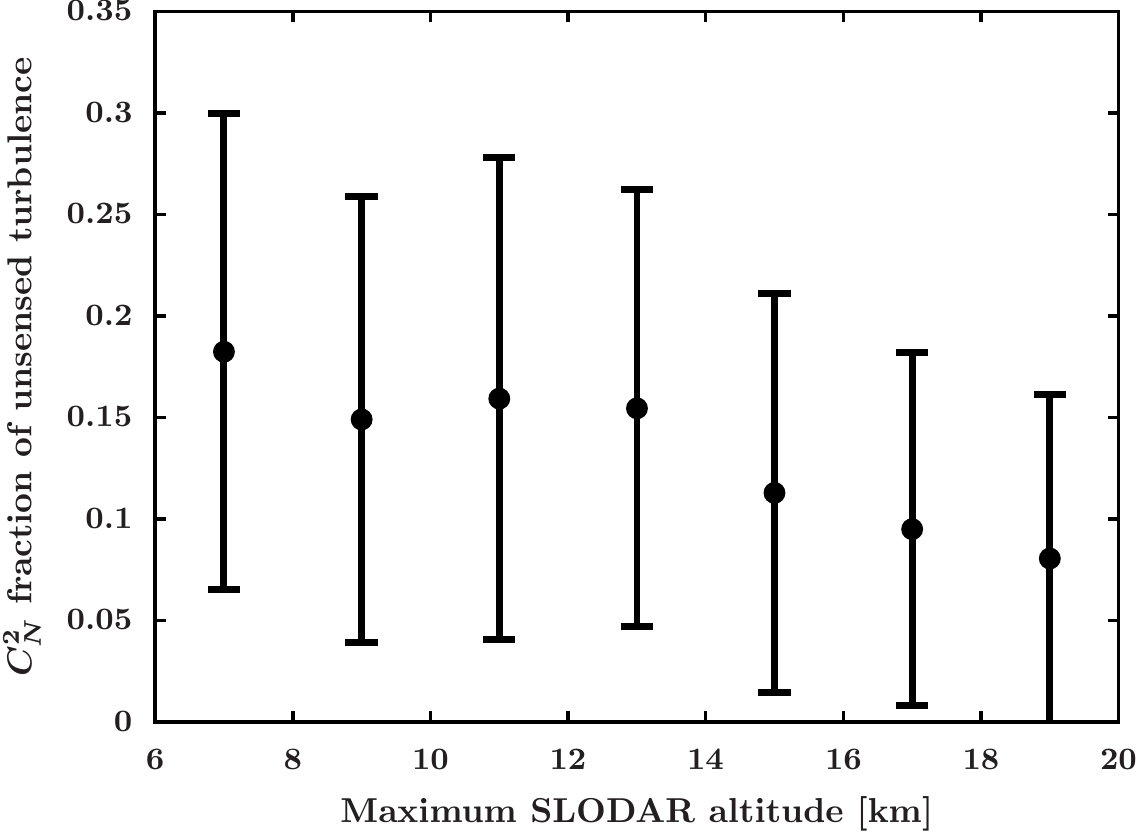}
\caption{Median $C_N^2$ fraction of the unsensed turbulence as a function of the maximum altitude $h_\text{max}$. The median fraction is computed in each 2\,km altitude bin from 6\,km to 20\,km, and the error bars represents standard deviations.} \label{fig:misscn2}
\end{figure}
Finally, we discuss more in-depth the unsensed turbulence. Currently, the unsensed turbulence is not taken into account in the tomography because the altitude of the unsensed turbulence can not be estimated by the SLODAR. However, this information should be useful for post-processing such as diagnosing the AO performance and PSF reconstruction.\par
\fref{fig:misscn2} shows the median $C_N^2$ fraction of the unsensed turbulence as a function of the maximum altitude $h_\text{max}$ that can be sensed by the cross-correlations. Although the fraction of the unsensed turbulence depends strongly on the vertical profile of $C_N^2$, the median value decreases with $h_\text{max}$. As $h_\text{max}=$6--8\,km, roughly 20\,\% of the turbulence are unsensed by the SLODAR. Even as the turbulence up to 18--20\,km are detected by the SLODAR, 8\,\% of the turbulence is unsensed. This unsensed turbulence directly affects performance of tomography and MOAO correction of RAVEN, and therefore, the unsensed turbulence should be considered in the evaluation of the on-sky MOAO performance of RAVEN. In ELTs, the effect of the unsensed turbulence gets smaller because the $h_\text{max}$ of the SLODAR increases with the telescope aperture diameter.\par
It should be noted that the unsensed turbulence estimation cannot be achieved by our SLODAR without NGS. In the case with LGSs, the $C_N^2$ unsensed turbulence detected by the auto-correlation depends on altitude due to the cone effect of LGSs. Some current and future WFAO systems have only LGSs (and tip/tilt NGS) and therefore more progress is needed to evaluate the unsensed turbulence with LGSs.
%
%
\section{Conclusion} \label{sec:conclusion}
In this paper, we present the fitted-SLODAR method to estimate the vertical profiles of $C_N^2$ and outer scale of the atmospheric turbulence using auto- and cross-correlations of slopes from multiple WFSs in WFAO systems. The analytical partial derivatives of slope correlations with respect to $r_0$ and $\mathcal{L}_0$ are developed and plugged in the form of the Jacobian to solve for the non-linear model-fit criterion minimisation for the SLODAR. Also, this SLODAR method can evaluate the unsensed turbulence, which cannot be sensed by the triangulation of the SLODAR. Finally, $C_N^2$ and $\mathcal{L}_0$ profiles at Maunakea are estimated by the fitted-SLODAR from on-sky telemetry data taken with multiple SH-WFSs in RAVEN during 12 nights of the RAVEN on-sky observations.\par
The main findings in this paper are as follows,
\begin{itemize}
\item The mean total seeing is $0.460^{\prime\prime}$ and it is better than the other result \citep{Els-09}. The $C_N^2$ fraction of the ground layer at $h<$1.5\,km is 54.3\,\%.
\item The mean profile of $C_N^2$ indicates that there is a strong turbulence at ground and weak turbulence at 8\,km. This profile has a good agreement with the mean $C_N^2$ profile estimated by the CFHT MASS-DIMM during the RAVEN observation, except for the ground layer. The $C_N^2$ difference in the ground layer suggests the ground layer depends strongly on the location. Also, the dome seeing may affect this difference in the ground layer. Our relatively weaker ground layer than that of the other comparisons contributes to the found good seeing condition during our observation runs.
\item The $C_N^2$ values at each altitude bin estimated by the SLODAR marginally correlates with those from the MASS-DIMM. However, the correlations have large dispersion due to the contamination from the other altitude bin especially at 2\,km bin. The correlation of the $C_N^2$ for the ground layer has a dispersion larger than that for the high altitudes ($h>1.5$\,km), and it suggests that the high altitude turbulence is relatively common for a large field of sky whereas the ground layer depends on the location.
\item The median $C_N^2$ fraction of the unsensed turbulence is 11.5\,\%. This fraction decreases with the maximum altitude $h_\text{max}$ that can be sensed by the cross-correlation: 18\,\% as $h_\text{max}=$6--8\,km and 7\,\% as $h_\text{max}=$18--20\,km.
\item The median value of the outer scale is 25.5\,m and the value is larger at higher altitude, which are consistent with the other results. On the other hand, the FWHM of on-sky PSF in H-band taken by RAVEN suggests outer scales larger than 30\,m, and it means that the estimates of $\mathcal{L}_0$ from the SLODAR may be biased towards 2--3 times of the telescope aperture due to the blindness of the SLDOAR to large outer scales. 
\end{itemize}
This new processed profiles are very useful to understand and improve the performance of RAVEN. In particular, the outer scale profile may have a large impact on the tip/tilt angular anisoplanatism over the field, therefore tomography. Also, the outer scale affects the estimation of $C_N^2$: To represent an optical path difference, the $C_N^2$ should be larger as $\mathcal{L}_0$ decreases. These effects will be more critical in tomographic system in ELTs, where the size of primary mirror is comparable to the typical outer scale size.\par
The possible improvement in our SLODAR is to profile the dome seeing. The dome seeing are suggested to have a very small $\mathcal{L}_0$ \citep{Guesalaga-16}, and usually the dome seeing is considered to largely contribute to the ground layer. The dome seeing can be estimated by assuming two turbulent modes with different outer scale at the ground, and this can easily be taken into account in our theoretical model.\par
Another point to be improved is the estimation of wind speed and directions. Recently, some algorithms for predictive atmospheric turbulence tomography in WFAO systems were proposed \citep{Correia-14,Ono-16}, which require monitoring the wind speed and direction at each altitude. The estimation of the wind information of the turbulence can be achieved using temporal correlation of measurements from multiple WFSs \cite{Ono-16}. Our theoretical model for slope correlations can include the wind speed and direction, and the wind speed and direction can be automatically estimated by fitting this theoretical model to the observed temporal correlation with different time delays.
%
%
\section*{Acknowledgements}
This work is supported of JSPS Grant-in-Aid for JSPS Fellows (15J02510) and A*MIDEX project (no. ANR-11-IDEX-0001-02) funded by the "Investissements d'Avenir" French Government program, managed by the French National Research Agency (ANR). Thanks to Olivier Martin for many discussions. Thanks to the staff members of Subaru Telescope for their support.



\bibliographystyle{mnras}



\appendix

\section{Partial deviation of the von Karman structure function}
Here, we present how to compute the partial deviation of the von Karman structure function in \eref{eq:vonkarman}. With respect to $r_0$, it is easily computed because $r_0$ is included in the first parenthesis as
\begin{equation}
\frac{\partial D_\phi(\rho)}{\partial r_0}=-\frac{5}{3}r_0^{-1}D_\phi(\rho).
\end{equation}
The partial deviation with respect to $\mathcal{L}_0$ is more complex than the case of $r_0$. To deal with deviation of the modified Bessel function of the second kind, we use the following expression,
\begin{equation}
\frac{\partial}{\partial x}\left\{x^\nu K_{\nu}(x)\right\}=-x^\nu K_{\nu-1}(x),
\end{equation}
where $x$ is $2\pi\rho/L_0$ in our case. The final formulation is given as
\begin{align}
\frac{\partial D_\phi(\rho)}{\partial L_0}&=\frac{5}{3}L_0^{-1}D_\phi-0.17253\frac{2^{1/6}}{\Gamma(5/6)}\left(\frac{L_0}{r_0}\right)^{5/3}\notag\\
&\hspace{5mm}\times\left(\frac{2\pi\rho}{L_0}\right)^{11/6}L_0^{-1}K_{-1/6}\left(\frac{2\pi\rho}{L_0}\right).
\end{align}


\bsp	
\label{lastpage}
\end{document}